\documentclass{PoS}

\usepackage{epsfig}

\def\beq{\begin{equation}}
\def\eeq{\end{equation}}
\def\bea{\begin{eqnarray}}
\def\eea{\end{eqnarray}}
\newcommand{\beqs}{\begin{subequations}}
\newcommand{\eeqs}{\end{subequations}}

\newcommand{\cref}[1]{Ref.~\cite{#1}}

\newcommand{\vev}[1]{\left<#1\right>}

\newcommand{\hh}{{\ensuremath{I{\kern-2.6pt h}}}}
\newcommand{\bhh}{{\ensuremath{\bar{I{\kern-2.6pt h}}}}}

\newcommand{\ka}{\kappa}
\newcommand{\la}{\lambda}
\newcommand{\Pb}{\bar{\Phi}}
\newcommand{\Hb}{\bar{H}}

\title{Double hybrid inflation and gravitational waves}

\ShortTitle{Double hybrid inflation and gravitational waves}

\author{\speaker{G. Lazarides}
\\
School of Electrical and Computer Engineering, Faculty of Engineering, 
Aristotle University of Thessaloniki, Thessaloniki 54124, Greece\\
E-mail: \email{lazaride@eng.auth.gr}}

\author{C. Panagiotakopoulos
\\
School of Rural and Surveying Engineering, Faculty of 
Engineering, Aristotle University of Thessaloniki, Thessaloniki 
54124, Greece\\
E-mail: \email{costapan@eng.auth.gr}}

\abstract{A double hybrid inflationary scenario in non-minimal 
supergravity which can predict values of the tensor-to-scalar 
ratio up to about $5\times 10^{-2}$ is presented. Larger values 
of this ratio would require unacceptably large 
running of the scalar spectral index. The underlying 
supersymmetric particle physics model possesses, for the chosen 
values of the parameters, practically two inflationary paths, 
the trivial and the semi-shifted one. The trivial path is 
stabilized by supergravity and supports a first stage of 
inflation with a limited number of e-foldings. The 
tensor-to-scalar ratio can become appreciable with the scalar 
spectral index remaining acceptable, as a result of the 
competition between the relatively mild supergravity and the 
strong radiative corrections to the inflationary potential. 
The additional number of e-foldings required for solving the 
puzzles of hot big bang cosmology are generated by a second 
stage of inflation along the semi-shifted path. This is 
possible only because the semi-shifted path is almost 
orthogonal to the trivial one and, thus, not affected by the 
strong radiative corrections on the trivial path and also 
because the supergravity effects remain mild. The model 
predicts the formation of an unstable network of open cosmic 
strings connecting monopoles to antimonopoles. This network 
decays to gravity waves well before recombination leading to 
possibly detectable signatures in future space-based laser 
interferometer gravitational-wave detectors.} 

\FullConference{18th International Conference From the Planck 
Scale to the Electroweak Scale \\
		25-29 May 2015\\
		Ioannina, Greece }

\begin{document}

\section{Introduction}

The recent results of BICEP2 \cite{bicep2} on the B-mode in the 
polarization of the cosmic microwave background radiation (CMBR) 
at degree angular scales indicate that inflationary scenarios 
may have to face a new challenge. Namely, they should be able 
to accommodate appreciable values of the tensor-to-scalar ratio 
$r$, since a B-mode could be due to the production of 
gravitational waves during inflation. Although $r$ seems 
\cite{criticism} to be smaller than initially claimed due to a 
possible underestimation of the foreground from Galactic 
polarized-dust emission, values of $r\sim 0.01$ cannot be 
excluded \cite{joint1,joint2}. The most recent joint analysis 
\cite{joint2} of the Planck and BICEP2 data yields $r\lesssim 
0.12$ at $95\%$ confidence level.

Supersymmetric (SUSY) hybrid inflation \cite{cop} -- for a 
review see e.g. Ref.~\cite{review} -- is 
undoubtedly one of the most promising inflationary scenarios.
In its simplest realization, though, it suffers from
some problems. The grand unified theory (GUT) gauge symmetry 
is spontaneously broken only at the end of inflation and, 
thus, if magnetic monopoles are predicted by this breaking, 
they are copiously produced \cite{smooth} leading to a 
cosmological catastrophe. Also, although accurate 
measurements \cite{planck15} imply that the scalar spectral 
index $n_{\rm s}$ is clearly lower than unity, this scenario 
gives \cite{senoguz} values very close to unity or even 
larger than unity within minimal supergravity (SUGRA).

These problems can be solved within a two stage variant 
of SUSY hybrid inflation with minimal SUGRA, known as 
standard-smooth hybrid inflation scenario \cite{stsmhi}. 
The cosmological scales exit the horizon during the 
first stage of inflation, which is of the standard hybrid 
type, along a trivial classically flat path on which the 
gauge group is unbroken. Restricting the number of 
e-foldings during this stage, we can achieve adequately 
low values of $n_{\rm s}$. The extra e-foldings needed 
for solving the horizon and flatness problems of hot big 
bang cosmology are generated by a second stage of 
inflation along a classically non-flat valley of minima, 
where the gauge group is broken. Consequently, magnetic 
monopoles are produced only at the end of the first 
inflationary stage, but are adequately diluted by the 
second stage. Note, in passing, that the idea of a two 
stage inflation has been used \cite{tetradis} in the past 
for solving the initial value problem of hybrid inflation.  

This scenario was realized within an extended SUSY 
Pati-Salam (PS) particle physics GUT model with only 
renormalizable interactions, which was constructed 
\cite{quasi} for a very different reason. Namely, the 
simplest SUSY PS model predicts \cite{hw} exact Yukawa 
unification \cite{als} and, with universal boundary 
conditions, yields unacceptable values of the $b$-quark 
mass. In the extended model, Yukawa unification is 
naturally and moderately violated and this problem is 
solved.

Here, we will show \cite{dinf} that a reduced version 
of this extended SUSY PS model based on the left-right 
symmetric gauge group $G_{\rm LR}=SU(3)_c\times 
SU(2)_{\rm L}\times SU(2)_{\rm R}\times U(1)_{B-L}$ 
can also yield a two stage inflationary scenario which 
can predict values of $r$ up to about $0.05$ together 
with acceptable values of $n_{\rm s}$. Larger values 
of $r$ would require unacceptably large running of 
$n_{\rm s}$. The first stage occurs along the 
trivial path, stabilized by SUGRA, and our present 
horizon undergoes a limited number of e-foldings. The 
obtained values of $r$ can be appreciable thanks to 
the presence of strong radiative and relatively mild 
SUGRA corrections to the inflationary potential. The 
second stage occurs on the so-called semi-shifted path 
\cite{semi}, where $U(1)_{B-L}$ is unbroken, and 
generates the extra e-foldings required. This is 
possible since the SUGRA corrections on the 
semi-shifted path also remain mild and this path, 
for the parameters chosen, is almost orthogonal to the 
trivial one and, thus, not affected by the strong 
radiative corrections on the trivial path.

After the termination of the first stage of inflation,
$SU(2)_{\rm R}$ breaks spontaneously to a $U(1)$ 
subgroup. This leads to the production of magnetic 
monopoles. The spontaneous breaking of a linear 
combination of this $U(1)$ and $U(1)_{B-L}$ at the 
end of the second inflationary stage leads to the 
formation of open cosmic strings connecting these 
monopoles to antimonopoles. At later times, the 
monopoles enter the post-inflationary horizon and the 
string-monopole system decays into gravity waves well 
before recombination without affecting the CMBR. The 
resulting gravity waves, however, may be measurable 
in the future space-based laser interferometer detectors.   

\section{The model in global SUSY}

The reduced version of the extended SUSY PS model of 
Ref.~\cite{quasi} which we will use here is based on 
the left-right symmetric gauge group $G_{\rm LR}=
SU(3)_c\times SU(2)_{\rm L}\times SU(2)_{\rm R}\times 
U(1)_{B-L}$, which is a subgroup of the PS gauge group. 
The superfields which are relevant for inflation are 
a conjugate pair of Higgs superfields $H$ and $\bar{H}$ 
in the $(1,1,2)_{1}$ and $(1,1,2)_{-1}$ representations 
of $G_{\rm LR}$, respectively, causing the breaking 
of $G_{\rm LR}$ to the standard model gauge group 
$G_{\rm SM}$, a gauge singlet $S$, and a pair of 
superfields $\Phi$, $\bar{\Phi}$ in the $(1,1,3)_{0}$ 
representation of $G_{\rm LR}$. The vacuum expectation 
value (VEV) $\vev{\Phi}$ of $\Phi$ breaks $G_{\rm LR}$ 
to $G_{\rm SM}\times U(1)_{B-L}$.

The superpotential relevant for inflation is
\beq
\label{eq:superpotential}
W=\kappa S\left(M^2-\Phi^2\right)-\gamma SH\bar{H}+
m\Phi \bar{\Phi} -\lambda\bar{\Phi}H\bar{H}.
\eeq
The parameters $M$, $m$ are superheavy masses, while 
$\ka$, $\gamma$, $\la$ are dimensionless constants.
All these parameters but one can be made real and 
positive by rephasing the superfields. For 
definiteness, we choose the remaining complex 
parameter to be real and positive too.

The resulting F-term scalar potential is 
\beq
V^0_F=|\kappa(M^2-\Phi^2)-\gamma H\Hb|^2+|m\Pb-2\ka S\Phi|^2
+|m\Phi-\la H\Hb|^2+|\gamma S+\la\Pb\,|^2\left(|H|^2+|\Hb|^2
\right).
\eeq
From $V^0_F$ and the vanishing of the D-terms, which implies 
that $\bar{H}^{*}=e^{i\theta}H$, one finds \cite{semi} two 
distinct continua of SUSY vacua:
\bea
\Phi=\Phi_{+},\quad\bar{H}^{*}=H,\quad |H|=
\sqrt{\frac{m\Phi_{+}}{\la}} \quad (\theta=0),\quad \Pb=S=0,
\\
\Phi=\Phi_{-},\quad\bar{H}^{*}=-H,\quad |H|=
\sqrt{\frac{-m\Phi_{-}}{\la}} \quad
(\theta=\pi), \quad \Pb=S=0,
\eea
where
\begin{equation}
\Phi_{\pm}\equiv\pm M \left(\sqrt{1+\left(\frac{\gamma m}
{2\ka\la M}\right)^2}\mp\frac{\gamma m}{2\ka\la M}\right). 
\label{eq:SUSYvacua1}
\end{equation}

The model generally possesses \cite{semi} three flat directions: 
$(i)$ The usual trivial path at 
\beq
\Phi=\Pb=H=\Hb=0 \quad{\rm with} \quad V^0_F=V_{\rm tr}\equiv\ka^2M^4,
\eeq 
where $G_{\rm LR}$ is unbroken. 
$(ii)$ The new shifted path at
\begin{equation}
\Phi=-\frac{\gamma m}{2\ka\la},\quad \Pb=-\frac{\gamma}{\la}\,
S,\quad H\Hb=\frac{\ka\gamma(M^2-\Phi^2)+\la m\Phi}
{\gamma^2+\la^2}
\end{equation}
with
\beq
V^0_F=V_{\rm nsh}\equiv\ka^2M^4\left(\frac{\la^2}
{\gamma^2+\la^2}\right)\left(1+\frac{\gamma^2m^2}
{4\ka^2\la^2M^2}\right)^2.
\end{equation}
This path supports new shifted hybrid inflation 
\cite{nshift} with $G_{\rm LR}$ broken to $G_{\rm SM}$. 
$(iii)$ The semi-shifted path, which exists only for 
$M^2>m^2/2\ka^2$, at
\beq
\Phi=\pm\,M\sqrt{1-\frac{m^2}{2\kappa^2M^2}},\quad
\Pb=\frac{2\ka\Phi}{m}\,S,\quad H=\Hb=0
\nonumber
\eeq
\beq
{\rm with}\quad V^0_F=V_{\rm ssh}\equiv
m^2M^2\left(1-\frac{m^2}{4\ka^2M^2}\right).
\nonumber
\eeq
It yields semi-shifted hybrid inflation \cite{semi} 
with $U(1)_{B-L}$ unbroken.

We take $M^2>m^2/2\ka^2$ and, thus, the semi-shifted 
path exists and, as one can show \cite{semi}, always 
lies lower than the trivial and the new shifted one.
We also take $\ka\sim 1$, $\gamma\ll \la\ll\kappa$, 
and $m\ll M$, so that the new shifted path (for $|S|<1$) 
essentially coincides with the trivial one and, thus, 
plays no independent role in our scheme.

\section{The first stage of inflation}

The first stage of inflation takes place along the trivial 
path, which, for large values of the canonically normalized 
inflaton, is stabilized by 
the SUGRA corrections. Although the number of e-foldings 
is limited, all the cosmological scales exit the horizon 
during this inflationary stage. Strong radiative and 
relatively mild SUGRA corrections to the inflationary 
potential then guarantee appreciable values of $r$ 
together with acceptable values of $n_{\rm s}$.

We adopt the K\"{a}hler potential
\begin{equation}
\label{kaehler}
K=-\ln\left(1-|S|^2\right)-\ln\left(1-|\bar{\Phi}|^2\right)+
|\Phi|^2+|H|^2+|\bar H|^2 
-2\ln\left(-\ln|Z_1|^2\right)+|Z_2|^2. 
\end{equation}
The two extra $G_{\rm LR}$ singlets $Z_1$ and $Z_2$ included 
in $K$ do not enter the superpotential. The resulting 
F-term potential in SUGRA is found to be
\begin{equation}
V_F=\left[\sum_i |W_{X_i}+K_{X_i}W|^2K_{X_i{X_i}^*}^{-1}-3|W|^2 
\right]e^K,
\end{equation}
where the sum is over all the fields $S,\,\Phi,\,\bar{\Phi},
\,H,\,\bar{H},\,Z_1,\,Z_2$ and a subscript $X_i$ 
denotes derivation with respect to $X_i$. The values of 
$Z_1$ and $Z_2$ are fixed \cite{pana} by anomalous 
D-terms. Note that the superfields $S,\,\bar{\Phi},
\,Z_1$ have no-scale type K\"{a}hler potentials which, 
in view of the relation
\begin{equation}
|K_{Z_1}|^2K_{Z_1{Z_1}^*}^{-1}=2,
\end{equation}
guarantee the exact flatness of the potential along the 
trivial path \cite{pana} and its approximate flatness 
on the semi-shifted one for $Z_2=0$. The relation 
\begin{equation}
|K_{Z_2}|^2K_{Z_2{Z_2}^*}^{-1}=|Z_2|^2\equiv\beta
\end{equation}
then implies -- cf. Ref.~\cite{pana} -- that, for $Z_2
\neq 0$, the complex inflatons $S$ and $\bar\Phi$ 
(approximately) for the two paths, respectively, acquire 
masses squared proportional to $\beta$.

Using the symmetries, we can rotate $S$ and $H$ on the 
real axis. The fields $\Phi,\,\bar{\Phi},\,\bar{H}$ remain 
in general complex. For simplicity, we restrict them on 
the real axis too. This will not 
influence our results in any essential way since these 
fields are anyway real in the vacuum and on all the flat 
directions. Also, we can show that, everywhere on the 
trivial and the semi-shifted inflationary paths, the 
mass-squared matrices of the imaginary parts of the fields do 
not mix with the mass-squared matrices of their real parts 
and, during both inflations, have positive eigenvalues in 
the directions perpendicular to these paths. So there is 
no instability in the direction of the imaginary parts of 
the fields which are orthogonal to these inflationary 
paths. Moreover, as we can prove, both the trivial and the 
semi-shifted inflationary paths are destabilized with the 
fields developing real values. 

The canonically normalized real scalar fields $\sigma,\,
\phi,\,\bar\phi,\,h,\,\bar{h}$ corresponding to the 
K\"{a}hler potential in Eq.~(\ref{kaehler}) are given 
by -- cf. Ref.~\cite{pana} -- 
\begin{equation}
S=\tanh\frac{\sigma}{\sqrt{2}}, \quad \Phi=\frac{\phi}
{\sqrt{2}}, \quad \bar{\Phi}=\tanh\frac{\bar{\phi}}
{\sqrt{2}},\quad H=\frac{h}{\sqrt{2}}, \quad\bar{H}=
\frac{\bar{h}}{\sqrt{2}}.
\end{equation}

We evaluate the potential $V_F$ with the factor $\exp{
\left[-2\ln\left(-\ln|Z_1|^2\right)+|Z_2|^2\right]}$ 
absorbed into redefined parameters $\kappa$, $\gamma$, 
$m$, and $\lambda$ and find
\begin{eqnarray}
 V_F  & = & \left[A_1^2 \cosh^2 \frac{\bar{\phi}}
{\sqrt{2}} -A_2^2 \sinh^2 \frac{\bar{\phi}}{\sqrt{2}}+
\beta A_3^2+A_4^2+A_5^2+\frac{1}{2}\left(h^2+\bar{h}^2 
\right)A_6^2 \right.
\nonumber
\\
& &+\left. \frac{1}{2} \left(\phi^2+ h^2+\bar{h}^2 
\right)A_3^2+\left(\sqrt{2} \phi A_5-2h\bar{h}A_6\right)
A_3\right]e^{\frac{1}{2}\left(\phi^2+h^2+\bar{h}^2 
\right)}.
\end{eqnarray}
Here   
\begin{equation}
\label{a1}
A_1=\kappa\left(M^2-\frac{\phi^2}{2}\right)-\frac{\gamma}{2} 
h\bar{h}, \quad A_2=m\frac{\phi}{\sqrt{2}}- 
\frac{\lambda}{2} h\bar{h}, 
\end{equation}
\begin{equation}
A_3=A_1 \sinh \frac{\sigma}{\sqrt{2}} \cosh \frac{\bar{\phi}}
{\sqrt{2}} + A_2 \cosh \frac{\sigma}{\sqrt{2}} 
\sinh \frac{\bar{\phi}}{\sqrt{2}}, 
\end{equation}
\begin{equation}
A_4=A_1 \sinh \frac{\sigma}{\sqrt{2}} \sinh \frac{\bar{\phi}}
{\sqrt{2}} + A_2 \cosh \frac{\sigma}{\sqrt{2}} \cosh 
\frac{\bar{\phi}}{\sqrt{2}}, 
\end{equation}
\begin{equation}
A_5=m \cosh \frac{\sigma}{\sqrt{2}} \sinh \frac{\bar{\phi}}
{\sqrt{2}} - {\sqrt{2}} \kappa \phi \sinh \frac{\sigma}
{\sqrt{2}} \cosh \frac{\bar{\phi}}{\sqrt{2}}, 
\end{equation}
and
\begin{equation}
A_6=\gamma \sinh \frac{\sigma}{\sqrt{2}} \cosh \frac{\bar{\phi}}
{\sqrt{2}} + \lambda  \cosh \frac{\sigma}{\sqrt{2}} \sinh 
\frac{\bar{\phi}}{\sqrt{2}}.
\end{equation}

On the trivial path ($\phi,\,\bar{\phi},\,h,\,\bar{h}=0$), 
$V_F$ becomes
\begin{equation}
\label{eq:VF1}
V_F=\kappa^2 M^4\left[1+\beta \sinh^2 \frac{\sigma}
{\sqrt{2}}\right].
\end{equation}
The mass-squared eigenvalues in the directions which are 
perpendicular to the trivial path, for $\sinh^2(
\sigma/\sqrt{2})\gg M^2/2$, are 
\begin{equation} 
m_{\phi}^2\simeq 4\kappa^2 \sinh^2\frac{\sigma}{\sqrt{2}}, \quad 
m_{\bar{\phi}}^2\simeq \kappa^2M^4\left[1+(1+\beta)\sinh^2
\frac{\sigma}{\sqrt{2}}\right],
\end{equation} 
and
\begin{equation}
\label{chi1} 
m^2_{{\chi}_1,{\chi}_2}=(\kappa M^2\mp\gamma)\left[\kappa M^2+
\left((1+\beta)\kappa M^2\mp\gamma\right)\sinh^2\frac{\sigma}
{\sqrt{2}}\right],
\end{equation} 
where $\chi_{1,2}=(h\pm\bar{h})/\sqrt{2}$. Note that, in 
particular, the mass-squared formulas in Eq.~(\ref{chi1}) hold 
for any value of $\sigma$. Thus, for $\gamma < \kappa M^2$, 
the trivial path is stable for large values of $|\sigma|$.
However, as $|\sigma|$ decreases, the eigenvalues and 
eigenstates of the $\phi-\bar{\phi}$ system change. When 
$\sinh^2(\sigma/\sqrt{2})\simeq M^2/2+m^2/2
\kappa^2 M^2$, one of these eigenvalues vanishes with 
$\bar{\phi}$ dominating the corresponding eigenstate.
As $\sinh^2(\sigma/\sqrt{2})\to M^2/2$, 
the eigenvalues become opposite to each other with 
$\phi$, $\bar{\phi}$ contributing equally to both the 
eigenstates. A further decrease of $\sinh^2(\sigma/
\sqrt{2})$ leads to the domination of the unstable 
eigenstate by $\phi$. Since $\phi$ must become nonzero to 
cancel the energy density $\kappa^2 M^4$ on the trivial 
path, we say that this path is destabilized at 
$\sigma_{\rm c}$ with
\begin{equation}
\label{critical1}
\sinh^2\frac{\sigma_{\rm c}}{\sqrt{2}}=\frac{M^2}{2}.
\end{equation}

To $V_F$ on the trivial path we add the dominant 
one-loop radiative corrections from the 
$N_{\phi}$-dimensional supermultiplet $\Phi$ 
($N_{\phi}=3$):
\begin{equation}
V_r^{\phi}=\kappa^2 M^4 \left(\frac{N_{\phi}\kappa^2}
{8\pi^2}\right) \ln \frac{2 \tanh^2\frac{\sigma}
{\sqrt{2}}}{M^2}.
\end{equation}
Note that the renormalization scale in these radiative 
corrections is chosen such that  $V_r^{\phi}$ vanishes 
at $|\sigma|=|\sigma_{\rm c}|$.

The full inflationary potential $V$ and its 
derivatives with respect to $\sigma$ (denoted by primes) 
are:
\begin{equation}
\frac{V}{\kappa^2 M^4}=1+\beta \sinh^2 \frac{\sigma}{\sqrt{2}}
+\frac{\delta_{\phi}}{4} \ln \frac{2 \tanh^2\frac{\sigma}
{\sqrt{2}}}{M^2} \equiv C(\sigma),
\end{equation}
\begin{equation}
\label{Vprime}
\frac{V^{\prime}}{\kappa^2 M^4}=\frac{1}{\sqrt{2}}
\sinh(\sqrt{2}\sigma)\left(\beta+\frac{\delta_{\phi}}
{\sinh^2(\sqrt{2}\sigma)}\right),
\end{equation}
\begin{equation}
\label{Vprimeprime}
\frac{V^{\prime \prime}}{\kappa^2 M^4}= \cosh(\sqrt{2}\sigma)
\left(\beta-\frac{\delta_{\phi}}{\sinh^2(\sqrt{2}\sigma)}\right),
\end{equation}
\begin{equation}
\frac{V^{\prime \prime \prime}}{\kappa^2 M^4}= \sqrt{2}
\sinh(\sqrt{2}\sigma)\left(\beta-\frac{\delta_{\phi}}
{\sinh^2(\sqrt{2}\sigma)}\right)
+\frac{2\sqrt{2}\delta_{\phi}}{\tanh^2(\sqrt{2}\sigma)
\sinh(\sqrt{2}\sigma)}
\end{equation}
with
\begin{equation}
\delta_{\phi}=\frac{N_{\phi}\kappa^2}{2\pi^2}.
\nonumber
\end{equation}
The usual slow-roll parameters for inflation are then 
\begin{equation}
\label{epsilon}
\epsilon=\frac{1}{2}\left(\frac{V^{\prime}}{\kappa^2 M^4}
\right)^2\frac{1}{C^2(\sigma)}, \quad \eta=\left(\frac{V^{
\prime \prime}}{\kappa^2 M^4}\right)\frac{1}{C(\sigma)},
\end{equation}
\begin{equation}
\xi=\left(\frac{V^{\prime}}{\kappa^2 M^4}\right)\left(
\frac {V^{\prime \prime \prime}}{\kappa^2 M^4}\right)
\frac{1}{C^2(\sigma)}=2\left|\tanh(\sqrt{2}\sigma)\right|
\eta\sqrt{\epsilon}
+\frac{4\delta_{\phi}\sqrt{\epsilon}}{C(\sigma)
\tanh^2(\sqrt{2}\sigma)\left|\sinh(\sqrt{2}\sigma)\right|}. 
\end{equation}
From these expressions, we evaluate $n_{\rm s}$, its 
running $\alpha_{\rm s}$, the tensor-to-scalar ratio $r$, 
and $V$: 
\begin{equation}
\label{ns}
n_{\rm s}=1+2\eta-6\epsilon, \quad 
\alpha_{\rm s}=16\eta\epsilon-24\epsilon^2-2\xi, 
\quad r=16\epsilon, \quad V=\frac{3\pi^2}{2}A_{\rm s} r,
\end{equation}
where $A_{\rm s}$ is the scalar power spectrum amplitude.

As a numerical example, take $\sigma_*=1.45$ at horizon 
exit of the pivot scale $k_*=0.05 \ \rm {Mpc}^{-1}$, 
$\kappa=1.7$, $\beta=0.022$, and $A_{\rm s}=2.215 \times 
10^{-9}$ at the same $k_*$ \cite{planck15}. We then find 
$M=3.493 \times 10^{-3}$, 
$C(\sigma_*)=2.2941$, $\epsilon=0.00188$, $\eta=-0.01389$, 
$n_{\rm s}=0.9609$, $r=0.0301$, and $\alpha_{\rm s}=-0.01674$.
So we can not only be consistent with the latest Planck 
data \cite{planck15}, but also accommodate large values of 
$r\sim {\rm few}\times 10^{-2}$. Note that large values of 
$r$ require relatively large values of $\epsilon$, which 
reduce $n_{\rm s}$ below unity, but not enough to make it 
compatible with the data. So large negative values of $\eta$ 
are needed, which requires that the parenthesis in the formula 
for $V^{\prime\prime}$ in Eq.~(\ref{Vprimeprime}) is dominated 
by the second term. A similar parenthesis appears in the 
formula for $V^{\prime}$ in Eq.~(\ref{Vprime}) too, but with 
the two terms added. So both these terms have to be appreciable 
with the second one being larger, which is possible only for 
large values of $\kappa$, which controls the radiative corrections 
on the trivial path. Inflation ends before the system reaches 
$\sigma_{\rm c}$ by violating the slow-roll conditions and the 
obtained number of e-foldings is limited due to the large 
values of $\epsilon$ and the fact that $\sigma_*\sim 1$.

\section{The second stage of inflation}

We choose, for the rest of the parameters, $m=1.827 \times 
10^{-5}$, $\lambda=0.1$, and $\gamma=10^{-6}$. Numerically, 
including also the D-terms from $H$ and $\bar{H}$, we find 
that there exist initial conditions for which, after the 
first stage of inflation, the energy density approaches 
$m^2M^2$ and ${\phi^2}\simeq 2M^2$, $h, \,\bar{h} \simeq 0$, 
$A_5 \simeq 0$ with $|\sigma| \ll 1$. So the system 
reaches the semi-shifted path and a second stage of 
inflation can take place. It is worth noticing that the 
initial values of the fields which lead to a double 
inflation scenario, although quite frequent, do not seem 
to form well-defined connected regions. In other words, 
the coupled system of differential equations exhibits a 
`chaotic' behavior, which means that a slight change of 
the initial conditions can possibly lead from a double to 
a single inflation scenario. A similar situation has been 
encountered \cite{initial} even in the simplest SUSY 
hybrid inflation scenario, where a slight change of 
initial conditions may ruin inflation. 

The potential $V_F$ on the semi-shifted path, for $M^2
\ll \beta$, is found to be
\begin{equation}
V_F\simeq m^2M^2\left[1+\beta \sinh^2 \frac{\bar{\phi}}
{\sqrt{2}}\right]. 
\end{equation}
Notice the striking similarity of this expression with the 
expression for $V_F$ on the trivial path in Eq.~(\ref{eq:VF1}). 
So the SUGRA corrections remain relatively mild on the 
semi-shifted path too. From $A_5 \simeq 0$, we find that the 
combination of $S$ and $\bar{\Phi}$ which is the complex 
inflaton during the second stage of inflation is 
\beq
\frac{mS+2\kappa <\Phi>\bar{\Phi}}
{\sqrt{m^2+4 \kappa^2 M^2}}\simeq \bar{\Phi},
\eeq
since $\bar{\Phi}$ contributes in this combination $2 \kappa 
M /m \simeq 650$ times more than $S$.

The mass eigenstates for the $h-\bar{h}$ system during the 
second stage of inflation are $\chi_{1,2}=(h\pm\bar{h})/\sqrt{2}$ 
with masses squared 
\beq
m^2_{{\chi}_1,{\chi}_2}=\left(\lambda\mp mM\right) \left[\left( 
\lambda \mp (1+\beta)mM\right)  \sinh^2 \frac{\bar{\phi}}
{\sqrt{2}}\mp mM\right].
\eeq
The eigenstate ${\chi}_1$ develops an instability terminating 
the semi-shifted valley of minima with the critical value 
$\bar{\phi}_{\rm c}$ of the real canonically normalized inflaton 
$\bar{\phi}$ being given by 
\beq
\sinh^2 \frac{\bar{\phi}_{\rm c}}{\sqrt{2}}=\frac{mM}{\lambda}.
\eeq

During the second stage of inflation, we include the 
dominant radiative corrections to the potential, which 
originate from the $N_{h}$-dimensional superfields $H$, 
$\bar{H}$ ($N_{h}=2$) and read as
\beq
V_r^{h} \simeq m^2 M^2 \left(\frac{N_{h}\lambda^2}{16\pi^2}
\right) \ln \frac{\lambda \tanh^2 \frac{\bar{\phi}}
{\sqrt{2}}}{mM}.
\eeq
The renormalization scale is chosen so that $V_r^{h}=0$ 
at $|\bar{\phi}|=|\bar{\phi}_{\rm c}|$. The radiative 
corrections from $\Phi$ are neglected since they are 
relatively very small. This is because $\Phi$ couples to 
the complex inflaton only through $S$ and the contribution 
of $S$ to this inflaton is severely suppressed. This is a 
very important property of the model resulting from the 
fact that, for the parameters chosen, the semi-shifted path 
is almost orthogonal to the trivial one. So the very strong 
radiative corrections on the trivial path, needed for 
accommodating appreciable values of $r$, do not affect the 
second stage of inflation. This is crucial since otherwise 
the semi-shifted path would be too steep to generate the 
extra e-foldings required. 

The number of e-foldings during the second stage of 
inflation between an initial $\bar{\phi}_{\rm {in}}$ and 
a final $\bar{\phi}_{\rm {f}}$ value of the inflaton 
$\bar{\phi}$ is $N(\bar{\phi}_{\rm {f}})-
N(\bar{\phi}_{\rm {in}})$, where
\beq
\label{ef}
N(\bar{\phi}) \simeq \frac{1}{2\beta \sqrt{1-(\delta_{h}/\beta)}}
\ln \frac{\cosh(\sqrt{2} \bar{\phi})+\sqrt{1-(\delta_{h}/\beta)}}
{\cosh(\sqrt{2} \bar{\phi})-\sqrt{1-(\delta_{h}/\beta)}}
\eeq
with $\delta_{h}=N_{h}{\lambda^2}/{4\pi^2}$. The termination of 
inflation is due to the radiative corrections and occurs at 
$\bar{\phi}=\bar{\phi}_{\rm f}$ ($|\bar{\phi}_{\rm f}|\gg|
\bar{\phi}_{\rm c}|$) with 
\beq
\cosh(\sqrt{2} \bar{\phi}_{\rm f}) \simeq 
\frac{\delta_{h}}{2}+\sqrt{1+\frac{\delta_{h}^2}{4}}.
\eeq

\begin{figure}[t]
\centerline{\epsfig{file=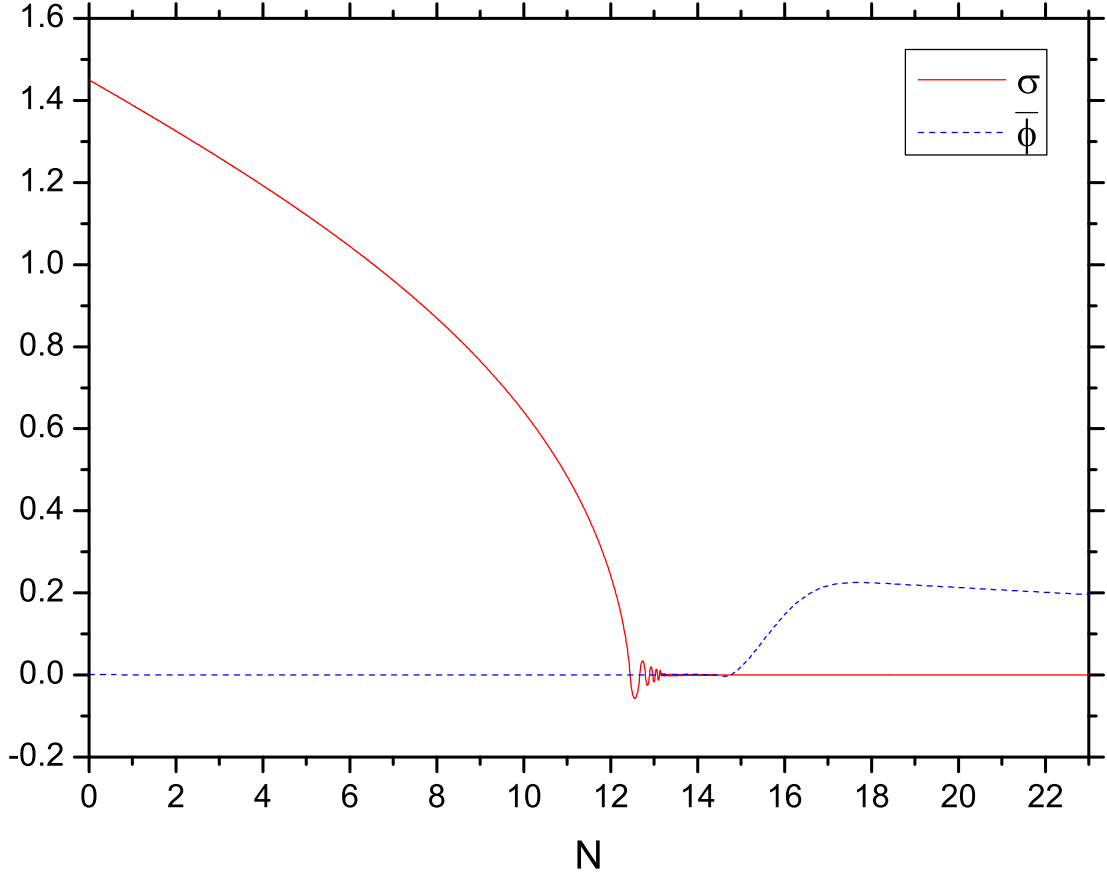,width=15cm}}
\caption{The evolution of $\sigma$ and $\bar\phi$ for the 
case with $r=0.0301$ versus the 
number $N$ of e-foldings after the horizon exit of the pivot 
scale $k_*$, where we take $\sigma=1.45$, $\bar\phi=10^{-3}$, 
$\phi=10^{-8}$, $h=10^{-4}$, $\bar{h}=1.01\times 10^{-4}$, 
and $d\sigma/dt=-1.1074\times 10^{-6}$.}
\label{fig1}
\end{figure}

Numerically, we find that, with the chosen values of 
the model parameters, the first stage gives rise to 
about 13 e-foldings. So another 38-39 e-foldings (for 
reheat temperature $T_{\rm r}=10^9~{\rm GeV}$) must 
be provided by the second stage of inflation, which 
requires that $|\bar{\phi}_{\rm in}|\simeq 0.23$ at 
the onset of this stage. This can indeed be satisfied in 
our numerical example as we have shown by extensive 
numerical calculations. It is actually remarkable that 
$|\bar\phi|$, which at the end of the first inflationary 
stage is extremely small, manages to attain values of 
the order of ${\rm few}\times 10^{-1}$ at the onset of 
the second stage.

To see this remarkable jump of $\bar\phi$ after the end of 
the first inflationary stage, we depict, in Fig.~\ref{fig1}, 
the evolution of the fields $\sigma$ and $\bar\phi$ as 
functions of the number of e-foldings $N$ starting from the 
point where the pivot scale $k_*=0.05~{\rm Mpc}^{-1}$ exits 
the horizon for a particular choice of initial conditions. 
Namely, we start with $\sigma=1.45$, $\bar\phi=10^{-3}$, 
$\phi=10^{-8}$, $h=10^{-4}$, and $\bar{h}=1.01\times 10^{-4}$. 
All the fields start with zero velocity except for $\sigma$ 
the initial velocity of which is taken to be $-1.1074\times 
10^{-6}$, which is its actual velocity on the trivial path 
determined numerically. We observe that $\sigma$ remains 
above its critical value 
for about $13$ e-foldings. Just before the end of the first 
inflationary stage, $\sigma$ oscillates four times around
zero with appreciable amplitude. When this amplitude falls 
below the critical value of $\sigma$, $\phi$ moves to its 
value on the semi-shifted path and $\bar\phi$ starts 
oscillating  slowly with variable amplitudes of order $M$. 
The size of $\bar\phi$ remains small for about $1.7$ 
e-foldings before starting its remarkable growth. This field 
acquires its largest value $\simeq 0.225$ at $N\simeq 17.7$, 
when the second inflationary stage has already started. 
For $N\gtrsim 20$, the evolution of $\bar\phi$ follows 
Eq.~(\ref{ef}) closely. 

\begin{figure}[t]
\centerline{\epsfig{file=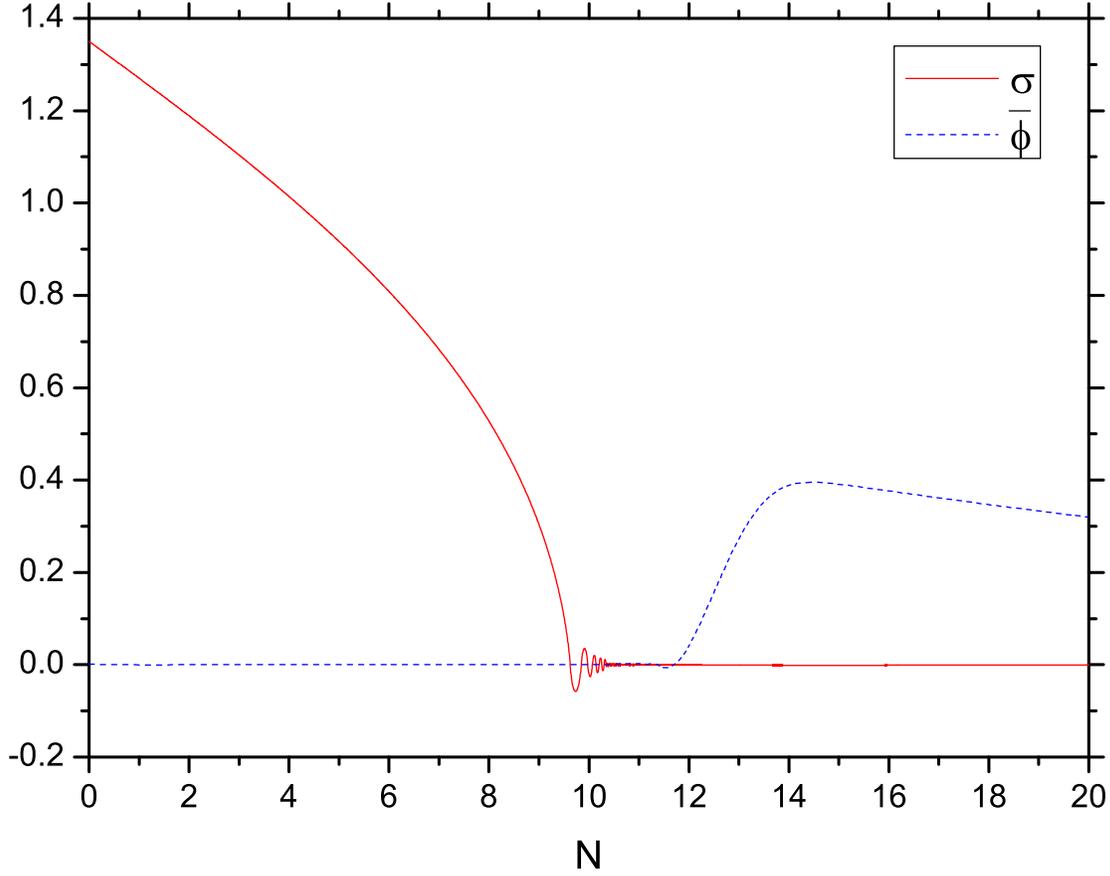,width=15cm}}
\caption{The evolution of $\sigma$ and $\bar\phi$ for 
the case with $r=0.0502$ versus the number $N$ of e-foldings 
after the horizon exit of the pivot scale $k_*$, where we take 
$\sigma=1.35$, $\bar\phi=10^{-3}$, $\phi=10^{-8}$, $h=9\times 
10^{-4}$, $\bar{h}=9.01\times 10^{-4}$, and 
$d\sigma/dt=-1.8523\times 10^{-6}$.}
\label{fig2}
\end{figure}

Allowing for a stronger running of $n_{\rm s}$, 
we can achieve larger values of $r$. For example, taking 
$\sigma_*=1.35$, $\kappa=1.75$, and $\beta=0.037$, we find 
that $M=3.891\times 10^{-3}$, $C(\sigma_*)=2.3479$, 
$\epsilon=0.00314$, $\eta=-0.00844$, $n_{\rm s}=0.9643$, 
$\alpha_{\rm s}=-0.03007$, and $r=0.0502$. In addition, we 
choose $m=3.891\times 10^{-5}$, $\lambda=0.1$, and 
$\gamma=10^{-6}$. In this case, the pivot scale suffers 
about $10$ e-foldings during the first inflationary stage 
and, consequently, approximately another 41-42 e-foldings 
must be provided by the second stage. This implies that 
$|\bar{\phi}_{\rm in}|$ must lie in the range 0.38-0.40, 
which is indeed feasible as we verified numerically. In 
Fig.~\ref{fig2}, we depict the evolution of $\sigma$ and 
$\bar\phi$ as functions of the number of e-foldings $N$ 
again starting from the point where the pivot scale $k_*$ 
exits the horizon, where we make a particular choice of 
initial conditions shown in the caption of the figure. 

Changing the values of the input parameters of the model, we can 
easily achieve successful solutions with smaller values of $r$. 
Note that there is no particular fine-tuning of the parameters 
required in our model. 

\section{Magnetic monopoles and cosmic strings}

Soon after the end of the first inflationary stage, the 
system settles on the semi-shifted path and $SU(2)_{\rm R}$ 
breaks spontaneously to a $U(1)$ subgroup by the nonzero 
value of $\Phi$. So magnetic monopoles are 
formed. An order of magnitude estimate of the mean 
monopole-antimonopole distance can be obtained as follows. 
At production, this distance is $p\, (2\kappa M)^{-1}$ as 
determined by the Higgs correlation length with 
$p\sim 1$ being a geometric factor. In the matter dominated 
era between the two inflationary stages, this distance is 
enhanced by a factor $\sim (\kappa^2 M^4/m^2M^2)^{1/3}$, 
where $\kappa^2 M^4$ and $m^2M^2$ are the classical 
potential energy densities on the trivial and the 
semi-shifted paths, respectively. The second inflationary 
stage stretches this distance by a factor $\exp{N_2}$ with 
$N_2$ being the corresponding number of e-foldings which 
is large but not huge -- cf. Ref.~\cite{intermediate}. 
During the damped inflaton oscillations, this distance 
increases by another factor $\sim (m^2M^2/c(T_{\rm r})
T_{\rm r}^4)^{1/3}$, where $T_{\rm r}\sim 10^{9}~{\rm GeV}$ 
is the reheat temperature and $c(T)=\pi^2g(T)/30$ with 
$g(T)$ being the effective number of massless degrees of 
freedom at cosmic temperature $T$. In the subsequent 
radiation dominated period, the monopole-antimonopole 
distance is multiplied by another factor $\sim T_{\rm r}
/T\sim (4c(T)/3)^{1/4}T_{\rm r}\sqrt{t}$, where $t$ is the 
cosmic time. So this distance, in the radiation dominated 
period, becomes
\beq
\sim \left(\frac{4}{3}\right)^{\frac{1}{4}} c(T_{\rm r})
^{-\frac{1}{3}}c(T)^{\frac{1}{4}}p\,
(2\kappa M)^{-1}e^{N_2}\left(\frac{\kappa^2M^4}
{T_{\rm r}^4}\right)^{\frac{1}{3}}T_{\rm r} t^{\frac{1}{2}}. 
\eeq
Equating this distance with the post-inflationary particle
horizon $\sim 2t$, we find the time $t_{\rm H}$ at which 
the monopoles enter this horizon:
\beq
t_{\rm H}\sim\frac{p^2}{8\sqrt{3}}\,
c(T_{\rm r})^{-\frac{2}{3}}c(T_{\rm H})^{\frac{1}{2}}e^{2N_2}
\left(\frac{M}
{\kappa\, T_{\rm r}}\right)^{\frac{2}{3}},
\label{tH} 
\eeq
with $T_{\rm H}$ being the cosmic temperature at time 
$t_{\rm H}$.

After the end of the second inflationary stage, the system 
settles in the SUSY vacuum and a linear combination of the 
$U(1)_{B-L}$ gauge symmetry and the unbroken $U(1)$ subgroup 
of $SU(2)_{\rm R}$ breaks 
spontaneously leading to the production of local cosmic 
strings. These strings, if they survived after recombination, 
could have a small contribution to the CMBR power spectrum 
which is parametrized \cite{bevis} by the dimensionless string 
tension $G\mu_{\rm s}$, where $G$ is Newton's constant and 
$\mu_{\rm s}$ is the string tension, i.e. the energy per 
unit length of the string. For local strings in the Abelian 
Higgs model in the Bogomol'nyi limit, the string tension is 
\cite{bevis}
\beq
\label{eq:stringtension}
\mu_{\rm s}=4\pi|\vev{H}|^2,
\eeq
where $\vev{H}$ is the VEV of $H$. Although the strings in our 
model are more complicated, we think that the above estimate 
is good enough for our purposes here. 

It is important that the strings, in our case, do not survive 
after recombination, but decay well before it. So they do not 
affect the CMBR. The reason is that they are open strings 
connecting monopoles to antimonopoles. This can be understood 
by simply realizing that the breaking of $SU(2)_{\rm R}\times 
U(1)_{B-L}$ to $U(1)_Y$ by $\vev{H}$ and $\vev{\bar H}$ is 
similar to the breaking of the electroweak gauge group and, 
thus, cannot lead to any topologically stable monopoles or 
strings -- for a more detailed explanation of this fact, see 
Ref.~\cite{trotta}. This breaking can only lead to the 
formation of topologically unstable dumbbell configurations 
\cite{dumbbell} consisting of an open string connecting a 
monopole to an antimonopole. 

At any time after their formation, the strings look like 
random walks with a step of the order of the particle horizon 
connecting monopoles to antimonopoles \cite{walsh} -- to 
describe the evolution of this string network, we will follow 
closely this reference. As argued in 
Ref.~\cite{walsh}, at all times before the entrance of 
monopoles into the horizon, there is of the order of one 
string segment per horizon and, thus, the ratio of the energy 
density $\rho_{\rm s}(t)$ of the string network to the total 
energy density $\rho_{\rm tot}(t)$ of the universe remains 
practically constant. At $t_{\rm H}$, there is approximately 
one monopole-antimonopole pair per horizon connected by an 
almost straight string segment of the size of the horizon. 
The energy density $\rho_{\rm s}(t_{\rm H})$ of the strings 
at $t_{\rm H}$ is then $\sim 3G\mu_{\rm s}/2t_{\rm H}^2$. After 
this time, more and more string segments enter the horizon, 
but the length of each segment remains constant. So the 
system of string segments behaves like pressureless matter. 
As a consequence, $\rho_{\rm s}(t)\sim 3G\mu_{\rm s}/
2(t_{\rm H}t^3)^{1/2}$ and the `relative string energy density'
\beq
\frac{\rho_{\rm s}(t)}{\rho_\gamma(t)}\sim 2G\mu_{\rm s}
\left(\frac{t}{t_{\rm H}}\right)^{\frac{1}{2}}
\label{strenergy}
\eeq
($\rho_\gamma(t)$ is the `photon' energy density) increases with 
time -- note that in a radiation dominated universe 
$\rho_\gamma(t)=\rho_{\rm tot}(t)$. The strings decay at cosmic 
time \cite{vilenkin}
\beq
t_{\rm d}\sim (\Gamma G\mu_{\rm s})^{-1}2t_{\rm H}
\eeq
with $\Gamma\sim 50$ by emitting gravity waves with energy 
density $\rho_{\rm gw}(t_{\rm d})$ at production given by
\beq 
\label{rhogwtd}
\frac{\rho_{\rm gw}(t_{\rm d})}{\rho_\gamma
(t_{\rm d})}\sim 2\left(\frac{2}{\Gamma}\right)
^{\frac{1}{2}}(G\mu_{\rm s})^{\frac{1}{2}}.
\eeq
Note that this is also the maximal relative string energy 
density. 

Taking the lower value of the number of e-foldings and $p=2$, 
Eq.~(\ref{tH}) gives, for the two numerical examples
presented, $t_{\rm H}\sim 4.8\times 10^{-7}~{\rm sec}$ and 
$1.04\times 10^{-4}~{\rm sec}$, respectively. Consequently, 
the strings enter
the horizon well before big bang nucleosynthesis. Their 
decay time is $t_{\rm d}\sim 6\times 10^{-2}~{\rm sec}$ and 
$5.5~{\rm sec}$, in the two cases, as one can infer from 
the corresponding dimensionless string tensions 
\beq
\label{Gmus}
G\mu_{\rm s}=\frac{|\vev{H}|^2}{2}\simeq \frac{mM}{2\lambda} 
\simeq 3.19 \times 10^{-7}\quad{\rm and}\quad 7.57\times 10^{-7}.
\eeq
So the strings decay well before recombination and, thus, do not 
affect the CMBR. Their maximal relative energy density is 
$\sim 2.3\times 10^{-4}$ and $3.5\times 10^{-4}$ for our two 
examples and, thus, the strings remain always subdominant. In 
particular, they do not disturb nucleosynthesis. 

Had the strings survived until now, an upper bound would have 
to be imposed on $G\mu_{\rm s}$ to keep their contribution to 
the CMBR power spectrum at an acceptable level. For the 
Abelian-Higgs field theory model, this bound is \cite{strings}
\beq
\label{Gmu}
G\mu_{\rm s}\lesssim 3.2\times 10^{-7}.
\eeq  
In our first numerical example, $G\mu_{\rm s}$ almost 
saturates this bound, but violates the recent bound 
\cite{nano} 
\beq
\label{Gmunano}
G\mu_{\rm s}\lesssim 3.3\times 10^{-8}
\eeq
from pulsar timing arrays, which also holds for
strings surviving until the present time. Our second numerical 
example violates both the bounds in Eqs.~(\ref{Gmu}) and 
(\ref{Gmunano}). Thus, both our examples are only possible 
because the strings decay sufficiently early to gravity waves.

The ratio of the energy density of the gravity waves produced 
by the strings to that of the photons at the present time $t_0$ 
can be found from Eq.~(\ref{rhogwtd}) to be -- cf. 
Ref.~\cite{maggiore} --
\beq 
\frac{\rho_{\rm gw}(t_0)}{\rho_\gamma(t_0)}\sim 2
\left(\frac{2}{\Gamma}\right)^{\frac{1}{2}}
(G\mu_{\rm s})^{\frac{1}{2}}\left(\frac{3.9}{10.75}
\right)^\frac{4}{3}
\eeq
and their present abundance is given by
\beq
\Omega_{\rm gw} h^2 (t_0)\sim \left(\frac{\rho_{\rm gw}(t_0)}
{\rho_\gamma(t_0)}\right)\left(\frac{\rho_\gamma(t_0)}
{\rho_{\rm c}(t_0)}\right)h_0^2,
\eeq
where $\rho_{\rm c}(t_0)$ is the present critical energy density 
of the universe and $h_0\simeq 0.7$ is the present value of the 
Hubble parameter in units of ${\rm km~sec^{-1}~Mpc^{-1}}$. As it 
turns out $\Omega_{\rm gw} h^2 (t_0)\sim 2.2\times 10^{-9}$ 
and $3.4\times 10^{-9}$ for our two numerical examples, 
respectively. The frequency $f(t_{\rm d})$ of these gravitational 
waves at production must be $\sim t_{\rm H}^{-1}$ since the length 
of the decaying strings is $\sim 2t_{\rm H}$ \cite{vilenkin}. The 
present value of this frequency is then 
\beq
f(t_0)\sim t_{\rm H}^{-1}\left(\frac{t_d}{t_{\rm eq}}\right)^
{\frac{1}{2}}\left(\frac{t_{\rm eq}}{t_0}\right)^{\frac{2}{3}},
\eeq  
where $t_{\rm eq}$ is the equidensity time at which matter starts
dominating the universe. For the two numerical examples, this 
frequency turns out to be $\sim 1.1\times 10^{-4}~{\rm Hz}$ and 
$4.7\times 10^{-6}~{\rm Hz}$, respectively. We see that these 
frequencies are too high to yield any restriction from CMBR 
considerations \cite{maggiore}. Also, they are well above 
the range probed by the pulsar timing array observations 
\cite{liu} and, thus, the recent bound \cite{nano} from pulsars 
does not apply to our case. However, the frequency of the 
gravity waves in our first numerical example lies marginally 
within the range to be probed by the future space-based laser 
interferometer gravitational-wave observatories such as 
eLISA/NGO \cite{lisa}. We conclude that the monopole-string 
system decays early enough without causing any trouble, 
but the gravity waves that it generates may be probed by 
future space-based laser interferometer observations.  
 
\section{Conclusions}

We considered a reduced version of the extended SUSY PS 
model of Ref.~\cite{quasi}, which was initially constructed 
for solving the $b$-quark-mass problem of the simplest SUSY 
PS model with universal boundary conditions. We find that 
this model can yield a two stage hybrid inflationary scenario 
predicting values of the tensor-to-scalar ratio of the 
order of ${\rm few}\times 10^{-2}$. For the values of the 
parameters chosen, the model in global SUSY possesses 
practically two classically flat directions: the trivial and 
the semi-shifted one. We have shown that the SUGRA corrections 
stabilize the trivial path, which can then support a first 
stage of inflation with a limited number of e-foldings. The 
obtained value of tensor-to-scalar ratio can be appreciable as 
a result of mild SUGRA corrections combined with strong 
radiative corrections to the inflationary potential, while the 
scalar spectral index remains acceptable.
 
The extra e-foldings required for solving the horizon and 
flatness problems of the standard hot big bang cosmological 
model are generated by a second stage of inflation along the 
semi-shifted path, where the gauge group $U(1)_{B-L}$ is 
unbroken. This is possible since the SURGA corrections to the 
potential on the semi-shifted path remain mild and this path 
is almost orthogonal to the trivial one and, thus, is not 
affected by the strong radiative corrections to the potential 
on the trivial path. 

After the termination of the first inflationary stage, magnetic 
monopoles are formed. Subsequently, at the end of the second 
stage of inflation, cosmic strings are produced connecting these 
monopoles to antimonopoles. At later times, the monopoles enter 
the horizon and the string-monopole system decays into gravity 
waves well before recombination without leaving any trace in the 
CMBR. The resulting gravity waves, however, may be measurable in 
the future.  

The baryon asymmetry of the universe can, in principle, be 
generated by non-thermal leptogenesis \cite{origin}. At 
reheating, the inflaton system decays into right-handed 
neutrino superfields, which subsequently decay out of 
equilibrium into light matter generating a primordial 
lepton asymmetry. This asymmetry is then partly turned into 
the observed baryon asymmetry of the universe by electroweak 
sphaleron effects. A detailed discussion of this mechanism 
within a particle physics model based on the same left-right 
symmetric gauge group as our present model, but with some 
differences, can be found in Ref.~\cite{armillis}.       

\def\ijmp#1#2#3{\emph{Int. Jour. Mod. Phys.}
{\bf #1}~(#2)~#3}
\def\plb#1#2#3{\emph{Phys. Lett. B }{\bf #1}~(#2)~#3}
\def\zpc#1#2#3{\emph{Z. Phys. C }{\bf #1}~(#2)~#3}
\def\prl#1#2#3{\emph{Phys. Rev. Lett.}
{\bf #1}~(#2)~#3}
\def\rmp#1#2#3{\emph{Rev. Mod. Phys.}
{\bf #1}~(#2)~#3}
\def\prep#1#2#3{\emph{Phys. Rep. }{\bf #1}~(#2)~#3}
\def\prd#1#2#3{\emph{Phys. Rev. D }{\bf #1}~(#2)~#3}
\def\npb#1#2#3{\emph{Nucl. Phys. }{\bf B#1}~(#2)~#3}
\def\np#1#2#3{\emph{Nucl. Phys. B }{\bf #1}~(#2)~#3}
\def\npps#1#2#3{\emph{Nucl. Phys. B (Proc. Sup.)}
{\bf #1}~(#2)~#3}
\def\mpl#1#2#3{\emph{Mod. Phys. Lett.}
{\bf #1}~(#2)~#3}
\def\arnps#1#2#3{\emph{Annu. Rev. Nucl. Part. Sci.}
{\bf #1}~(#2)~#3}
\def\sjnp#1#2#3{\emph{Sov. J. Nucl. Phys.}
{\bf #1}~(#2)~#3}
\def\jetp#1#2#3{\emph{JETP Lett. }{\bf #1}~(#2)~#3}
\def\app#1#2#3{\emph{Acta Phys. Polon.}
{\bf #1}~(#2)~#3}
\def\rnc#1#2#3{\emph{Riv. Nuovo Cim.}
{\bf #1}~(#2)~#3}
\def\ap#1#2#3{\emph{Ann. Phys. }{\bf #1}~(#2)~#3}
\def\ptp#1#2#3{\emph{Prog. Theor. Phys.}
{\bf #1}~(#2~#3)}
\def\apjl#1#2#3{\emph{Astrophys. J. Lett.}
{\bf #1}~(#2)~#3}
\def\apjs#1#2#3{\emph{Astrophys. J. Suppl.}
{\bf #1}~(#2)~#3}
\def\n#1#2#3{\emph{Nature }{\bf #1}~(#2)~#3}
\def\apj#1#2#3{\emph{Astrophys. J.}
{\bf #1},~#3~(#2)}
\def\anj#1#2#3{\emph{Astron. J. }{\bf #1},~#3~(#2)}
\def\mnras#1#2#3{\emph{MNRAS }{\bf #1},~#3~(#2)}
\def\grg#1#2#3{\emph{Gen. Rel. Grav.}
{\bf #1}~(#2)~#3}
\def\s#1#2#3{\emph{Science }{\bf #1}~(#2)~#3}
\def\baas#1#2#3{\emph{Bull. Am. Astron. Soc.}
{\bf #1}~(#2)~#3}
\def\ibid#1#2#3{\emph{ibid. }{\bf #1}~(#2)~#3}
\def\cpc#1#2#3{\emph{Comput. Phys. Commun.}
{\bf #1}~(#2)~#3}
\def\astp#1#2#3{\emph{Astropart. Phys.}
{\bf #1}~(#2)~#3}
\def\epjc#1#2#3{\emph{Eur. Phys. J. C}
{\bf #1}~(#2)~#3}
\def\nima#1#2#3{\emph{Nucl. Instrum. Meth. A}
{\bf #1}~(#2)~#3}
\def\jhep#1#2#3{\emph{J. High Energy Phys.}
{\bf #1}~(#2)~#3}
\def\jcap#1#2#3{\emph{J. Cosmol. Astropart. Phys.}
{\bf #1}~(#2)~#3}
\def\lnp#1#2#3{\emph{Lect. Notes Phys.}
{\bf #1}~(#2)~#3}
\def\jpcs#1#2#3{\emph{J. Phys. Conf. Ser.}
{\bf #1}~(#2)~#3}
\def\aap#1#2#3{\emph{Astron. Astrophys.}
{\bf #1}~(#2)~#3}
\def\mpla#1#2#3{\emph{Mod. Phys. Lett. A}
{\bf #1}~(#2)~#3}

\end{document}